\documentclass[twocolumn,showpacs,preprintnumbers,aps,amsmath]{revtex4}
\usepackage{amsmath,amssymb}
\usepackage{latexsym}
\usepackage[dvips]{graphicx}

\begin{document}

\preprint{EPHOU 07-003 }

\title{Integrable Cosmological Models From Higher Dimensional Einstein Equations}

\author{Masakazu Sano and Hisao Suzuki}

\affiliation{ Department of Physics,  Hokkaido University \\ Sapporo, Hokkaido 060-0810 Japan}

\begin{abstract}
We consider the cosmological models  for the higher dimensional spacetime which includes the curvatures of our space as well as the curvatures of the internal space.  We find that the condition for the integrability of the cosmological equations is that
the total space-time dimensions are $D=10$ or $D=11$ which is exactly the conditions for superstrings or M-theory.
We obtain analytic solutions with generic initial conditions in the four dimensional Einstein frame  and study the accelerating universe when both our space and the internal space have negative curvatures. 
\end{abstract}


\maketitle

\section{Introduction}

There has been much attention to the understanding cosmology from the superstring or M-theory.
The recent observation of Type Ia supernovae and CMB measurement by WMAP indicates our universe is an accelerating universe. In String/M-Theory, there is the no-go theorem \cite{bio1} which is an obstacle to 
realize the accelerating universe. The no-go theorem indicates that the warped compactification with the static internal space does not give rise to the four-dimensional de Sitter spacetime 
under some assumptions. This implies that the warped compactification can not lead the accelerating universe with the static internal space.

One way to avoid the no-go theorem is to employ the time-dependent internal space \cite{bio2}. 
pace. If the curvature of the internal space is negative in the four-dimensional Einstein frame, it has been shown that the effective Lagrangian has positive potentials \cite{bio3}-\cite{bio7}.  This positive potential gives rise to the acceleration of 
the four-dimensional spacetime in the four-dimensional Einstein frame. In the four-dimensional Einstein frame, the scale factor of the internal space appears as a scalar field with potentials which come from the curvature of the internal s
S-brane solutions also lead to accelerating solutions \cite{bio8}-\cite{bio14}. The flux field has a role of a positive potential and 
contributes to the acceleration of the four-dimensional spacetime. Scalar perturbations of scale factors have 
been shown in \cite{bio4} that the eternally accelerating universe is realized in the eleven-dimensional spacetime 
by the external and internal spaces which possess a negative curvature. Recently,  
it has been also shown by scalar perturbations of scale factors that the eternally accelerating universe occurs in the ten-dimensional spacetime \cite{bio5}.
The fixed point analysis \cite{bio29} also showed that the eternal acceleration is realized if two spaces possess a negative curvature. 

In general, it is difficult to solve the Einstein equations exactly because of the non-linearity of the Einstein equations.  Therefore, most of the analytic solutions are special solutions with particular initial conditions.  However, it is desirable to find integrable models for the analysis of the initial conditions for our universe.  Up to now, few integrable models have been found. In $p$-brane and cosmological solutions which were inspired with String/M-theory, it is known that there 
exist a few classes of models whose solutions can be reduced to the Liouville or Toda type \cite{bio15}-\cite{bio21}. 
The Toda equation is integrable and provides exact solutions for us. 
In \cite{bio15}-\cite{bio21}, the metric has two spatial parts whose curvature is flat for all spaces 
or for one of two spaces.   
However, if both spaces have curvatures, it is very difficult to solve even vacuum Einstein equations exactly. When our universe starts from the quantum era, it is natural to expect that we have spatial curvatures whose values are comparable with the curvature to the direction of time.  The  curvatures with respect to the spatial directions can be regarded as potential energies whereas the time dependence can be regarded as the kinetic energy of the effective actions. If the universe started from quantum fluctuations, it is natural to expect that the order of the potential energies are the same as the order of the kinetic energy. Therefore, it is highly desirable to find integrable models with spatial curvatures both for our space and the internal space.

The accelerating universe occurs when two spatial parts have the negative curvature as shown 
in \cite{bio4}-\cite{bio5}. This analysis performed by the perturbation of scale factors. 
In \cite{bio29} the fixed point analysis was performed and also showed that the eternally accelerating universe 
occurs with two spatial parts whose curvature is negative. 
In this paper,  we will try to solve $D$-dimensional vacuum Einstein equations with two homogeneous spaces exactly.

In order to find integrable cosmological models, we will adopt a method for solving  Einstein equations analyzed in \cite{bio22}. In the method used in \cite{bio22}, the Einstein equations can be reduced to an analytically mechanical problem with one gauge degree of freedom. The gauge degree of freedom originates from the choice of the time variable. By using time variables which include scalar fields, the condition of the integrability has been classified. We will review this method in the next section and find that we are able to obtain general solutions for those vacuum Einstein equations. Actually, it will turn out that the classification of \cite{bio22} was not complete and we will show that a new type of integrable models is useful for our analysis.

We will start with the generic space ansatz with total dimensions $D$ and our spatial dimensions $d$. 
Strangely enough,  we will find that the condition for the integrability is that the total dimension is 
ten or eleven, which is exactly correspond to the consistency conditions of superstrings or M-theory. If the integrable condition is satisfied there is an additional conserved quantity aside from the Hamiltonian constraint. Therefore the system has two conserved quantities for two dynamical variables 
and then the system reduces to the integrable case. 

In \cite{bio26}-\cite{bio28}, the relation between the integrable condition and the conserved quantity was investigated from the Hamiltonian 
viewpoint.  The same integrability condition was obtained and it was shown that there exist 
the conserved quantity under the integrability condition. 

The integrable system does not necessarily have the simple analytic solutions \cite{bio26}-\cite{bio28}. 
In this paper, it will be  shown that we are able to derive the analytic solutions 
by the particular choices of the time variable in $D=10$ and $D=11$. 
It is very important to note that the time variable used in this paper 
easily realizes the analytic solutions. 
The other choices of the time variable make it difficult to solve the equations of motion.

We investigated the cosmological behavior in $D=10$ because this case includes the four-dimensional spacetime and the six-dimensional internal space.
It was found that the accelerating universe occurs if two spaces have a negative curvature.  

This paper is organized as follows. In the next section, we will construct the effective action which has 
two potential terms arising from curvatures of the two spaces. We will also show that the system is integrable when the total dimensions of the space-time is ten or eleven. 
 In section three, we will solve Einstein equations completely and 
general solutions will be given. 
 In section four, we will consider the cosmological property of 
the case of ($D=10,\,d=3$) in detail and the 
analytic solutions representing the accelerating universe when two spatial parts have negative curvature. 
The asymptotic behavior of the metric is also discussed.  
Finally, we conclude with a discussion of our results and some implications.

\section{Effective Action and conditions for integrability}
Let us consider the $D$-dimensional spacetime constructed with two spatial parts which have  curvatures. For simplicity, we will consider the vacuum Einstein Equations.  We consider that the $D$-dimensional space consists of two homogeneous spaces whose sizes depend on time.
Namely, the metric ansatz for the $D$-dimensional spacetime is given by, 
\begin{align}
ds_{D}^{2}&=G_{MN}(X )\,dX^{M}dX^{N} \notag \\
&=-e^{2\,\widetilde{n}(t)}dt^{2}+e^{2\,\widetilde{A}(t)} g_{ij}(x)\,dx^{i}dx^{j} \notag \\
&\qquad \qquad \qquad \qquad +e^{2\, \alpha (t)}g_{AB}(y)\,dy^{A}dy^{B},  \label{a1}
\end{align} 
where 
\begin{equation}
\begin{cases}
i,\,j=1,\,2,\,\cdots ,\,d \\
A,\,B=d+1,\,d+2,\,\cdots ,\,D-1 \\
M,\,N=0,\,1,\,\cdots ,\,D-1. \\
\end{cases} \notag 
\end{equation}
$e^{2\,\widetilde{A}(t)}$ and $e^{2\,\alpha(t)}$ are scale factors of the $d$- and $(D-d-1)$-dimensional spaces whose metrics are $g_{ij}(x)\,dx^{i}dx^{j}$ and $g_{AB}(y)\,dy^{A}dy^{B}$. respectively. 
These scale factors depend on time variable $t$.

We assume that  both two spatial manifolds are homogeneous spaces (Einstein spaces);   
\begin{align}
{}^{(d)} \widetilde{R}_{ij}(g(x))&=(d-1)\,k_{(d)}\,g_{ij}(x), \label{a2} \\
{}^{(D-d-1)} \widetilde{R}_{AB}(g(y))&=(D-d-2)\,k'_{(D-d-1)}\,g_{AB}(y), \label{a3}
\end{align}
where $k_{(d)}$ and $k'_{(D-d-1)}$ represent the curvature of Einstein spaces.  For our physical $d$-dimensional space, we assume 
homogeneous and isotropic space. However we do not need the explicit representation of $g_{ij}(x)$ and $g_{AB}(y)$ to derive 
the effective action. 

 The ($d+1$)-dimensional Einstein frame is realized by the following 
conformal transformations; 
\begin{equation}
\begin{split}
&\widetilde{A}(t) \longrightarrow A(t)-\frac{1}{d-1}\,(D-d-1)\,\alpha (t),  \\
&\widetilde{n}(t)\longrightarrow n(t)-\frac{1}{d-1}\,(D-d-1)\,\alpha (t).
\end{split} \label{a8}
\end{equation}
In fact, the $D$-dimensional Einstein$-$Hilbert Lagrangian can be written as 
\begin{equation}
\begin{split}
 \sqrt{-G}\,R&=\sqrt{-g_{d+1}(x)\,g_{D-d-1}(y)}\,e^{d\,\widetilde{A}+\widetilde{n}+(D-d-1)\,\alpha}\,R \\
\longrightarrow & \sqrt{-g_{d+1}(x)\,g_{D-d-1}(y)}\,e^{d\,A+n}\,(\,{}^{(d+1)}R+\cdots \,).  
\end{split} \notag 
\end{equation}
Under the conformal 
transformation (\ref{a8}), we obtain an effective action in the ($d+1$)-dimensional Einstein frame \cite{bio3};
\begin{equation}
\begin{split}
&\frac{1}{d(d-1)\,V_{D-1}}\int d^{D}X\,\sqrt{-G}\,R \\ 
\longrightarrow   &\int dt  \,
\Biggl[ e^{dA-n}\,\Bigl\{ -\dot{A}^{2}+X^{2}\,\dot{\alpha}^{2} \Bigr\} \\
&-e^{dA+n}\Bigl\{ \, -k_{(d)}\,e^{-2A}-\widetilde{k}_{(D-d-1)}\,e^{-\frac{2(D-2)}{d-1}\,\alpha} \, \Bigr\} \Biggr], 
\end{split} \label{a9}
\end{equation}
where $V_{D-1}\equiv \int d^{D-1}X\,\sqrt{-\det g_{ij}(x)}\,\sqrt{\det g_{AB}(y)}$ and 
\begin{equation}
\begin{split}
X^{2}&=\frac{(D-d-1)(D-2)}{d(d-1)^{2}}, \\
\widetilde{k}_{(D-d-1)}&=\frac{(D-d-1)(D-d-2)}{d(d-1)}\,k'_{(D-d-1)}.
\end{split} \label{a10}
\end{equation} 
The effective action (\ref{a9}) shows that the scale factor of the internal space 
appears as a scalar field with the potential terms. The
two effective potentials are generated by the curvature of 
Einstein spaces which can be easily seen from the fact that the potential terms are directly proportional to the curvatures.

We would like to solve equations of motion derived from the ($d+1$)-dimensional effective action (\ref{a9}). 
In general, Einstein equations are highly non-linear equations and difficult to be solved exactly. The effective action (\ref{a9}) indeed results in non-linear equations.  A way to analyze the system is to utilize a gauge degree of freedom \cite{bio24}. Let us recall that 
we can choose a time variable via the lapse function $e^{n(t)}$ which is a non-dynamical quantity \cite{bio24}.  This gauge degree of freedom represents the invariance under the coordinate transformation of time. Because of this gauge degree of freedom, we can freely choose the gauge to solve the system. 
We will take the lapse function as
\begin{equation}
e^{n(t)}=e^{p\,\alpha(t)+q\,A(t)}, \label{a11} 
\end{equation}
where $p$ and $q$ are any real numbers. 
We would like to look for the more convenient transformation of two dynamical variables.

We use the following transformation between ($A,\,\alpha$) and ($U_{+},\,U_{-}$),  
\begin{align}
e^{2A}&=U_{+}^{\frac{M_{1}+1}{d-1}}\,U_{-}^{\frac{M_{2}+1}{d-1}}, \notag \\
e^{2\alpha}&=U_{+}^{\frac{(M_{1}+1)}{(d-1)\,X}}\,U_{-}^{-\frac{(M_{2}+1)}{(d-1)\,X}}, \label{a14} \\
e^{2n}&=U_{+}^{-2+\frac{d\,(M_{1}+1)}{d-1}}\,U_{-}^{-2+\frac{d\,(M_{2}+1)}{d-1}}, \notag 
\end{align}
where we have defined 
\begin{equation}
\begin{split}
M_{1}=\frac{(d-2+q)\,X+p}{(d-q)\,X-p}, \\
M_{2}=\frac{(d-2+q)\,X-p}{(d-q)\,X+p}. \label{a13}
\end{split}
\end{equation}
By using these variables, we can re-write the effective action (\ref{a9}) as \cite{bio22}
\begin{align}
&\mathcal{L}_{\text{eff}}=-\frac{(M_{1}+1)(M_{2}+1)}{(d-1)^{2}}\,\dot{U}_{+}\dot{U}_{-} 
+k_{(d)}\,U_{+}^{M_{1}}U_{-}^{M_{2}} \notag \\
&+\widetilde{k}_{(D-d-1)}\,U_{+}^{-1+\frac{M_{1}+1}{d-1}\Bigl[ d-\frac{D-2}{
(d-1)X} \Bigr]}U_{-}^{-1+\frac{M_{2}+1}{d-1}\Bigl[d+\frac{D-2}{
(d-1)X} \Bigr]}. \label{a12}
\end{align}
The effective action $\int dt \mathcal{L}_{\text{eff}}(M_{1},\,M_{2})$ can be transformed to the action  of some other parameters $N_{1},N_{2}$,
$\int d\xi \mathcal{L}_{\text{eff}}(N_{1},\,N_{2})$ by the change of time coordinates, $dt=N(\xi)\, d\xi$. 
This is realized by following transformations; 
\begin{equation}
\begin{split}
&U_{+}=V_{+}^{\frac{N_{1}+1}{M_{1}+1}}, \quad
U_{-}=V_{-}^{\frac{N_{2}+1}{M_{2}+1}}, \\
&N=V_{+}^{-1+\frac{N_{1}+1}{M_{1}+1}}V_{-}^{-1+\frac{N_{2}+1}{M_{2}+1}}. 
\end{split} \label{a15}
\end{equation} 
These transformations preserve the form of the effective action (\ref{a12}). 
It is possible to connect a solution in some parameters ($M_{1},\,M_{2}$) to 
many other solutions by above transformations (\ref{a15}). Because of this gauge degree of freedom, we can solve the Einstein equations 
with a particular choice of the parameters.

In \cite{bio22}, $M_{1}=M_{2}=0$ was considered. 
In our cases, this condition leads the following potential 
 \begin{align}
 W&=-k_{(d)} \notag \\
&-\widetilde{k}_{(D-d-1)}U_{+}^{-1+\frac{1}{d-1}\Bigl[ d-\frac{D-2}{
(d-1)X} \Bigr]}U_{-}^{-1+\frac{1}{d-1}\Bigl[ d+\frac{D-2}{
(d-1)X} \Bigr]}. \notag 
\end{align}
If the form of the potential becomes $W=-k_{(d)}
-\widetilde{k}_{(D-d-1)}\,U_{+}$ or $W=-k_{(d)}
-\widetilde{k}_{(D-d-1)}\,U_{-}$, Einstein equations are soluble as shown in \cite{bio22}. 
A model with this type of the potential also studied in \cite{bio25}.  
But the above potential cannot take such that potential 
because the total dimension has to be $D=1$ in order to satisfy $-1+\frac{1}{d-1}\,\Bigl[\, d\pm \frac{D-2}{
(d-1)X} \,\Bigr]=0$. Therefore our model is not correspond to the model considered in \cite{bio22}. 

A convenient choice of the parameters is 
$M_{1}=M_{2}=1$
  which implies 
\begin{align}
&\mathcal{L}_{\text{eff}}=-\frac{4}{(d-1)^{2}}\,\dot{U}_{+}\dot{U}_{-} 
+k_{(d)}\,U_{+}\,U_{-} \notag \\
&+\widetilde{k}_{(D-d-1)}\,U_{+}^{-1+\frac{2}{d-1}\,\Bigl[\, d-\frac{D-2}{
(d-1)X} \,\Bigr]}U_{-}^{-1+\frac{2}{d-1}\,\Bigl[\, d+\frac{D-2}{
(d-1)X} \,\Bigr]}. \label{a16}
\end{align}
The above effective Lagrangian shows the second term is an interaction similar to a harmonic oscillator and third term 
represents a non-linear interaction which is an obstacle to solve equations of motion. If the power of $U_{+}$ or 
$U_{-}$ is simplified, it is possible to solve the equations analytically. To perform this 
procedure, we will impose a condition in which the non-linear term in the effective action does not depend on $U_+$;
\begin{equation}
 -1+\frac{2}{d-1}\,\Bigl[\, d-\frac{D-2}{(d-1)X}\Bigr]=0 ,\label{condition}
\end{equation}
where $X$ was defined in (\ref{a10}).
We will see below that the system is integrable if this condition is satisfied.
Before considering the integrability, let us solve the condition (\ref{condition}). The condition (\ref{condition}) can be rewritten as
\begin{equation}
(d-1)(D-d-5)=4.
\end{equation}
From this, we can immediately derive the condition of the integrability $D$ and $d$ as follows 
\begin{equation}
\begin{cases}
D=10,\qquad d=3 \\
D=11,\qquad d=2,\,5,
\end{cases} \label{a17}
\end{equation} 
where (\ref{a10}) was used. 
Note that $D=10$ is the critical dimension of the superstring theories and $D=11$ is the dimension of the M-theory! 
Moreover, $d=3$ means our spacetime is four dimensions.  Therefore, we have integrable cosmological models for a realistic setup.

We are going to show that the system is integrable if the condition (\ref{condition}) is satisfied.
By using the condition,  we can rewrite the effective Lagrangian and the Hamiltonian constraint as
\begin{align}
\mathcal{L}_{\text{eff}}&=-\frac{4}{(d-1)^{2}}\,\dot{U}_{+}\dot{U}_{-} 
+k_{(d)}\,U_{+}U_{-} \notag \\
& \qquad \qquad \qquad \qquad +\widetilde{k}_{(D-d-1)}\,U_{-}^{2(d+1)/(d-1)}, \label{a18} \\
H&=0=-\frac{4}{(d-1)^{2}}\,\dot{U}_{+}\dot{U}_{-} 
-k_{(d)}\,U_{+}U_{-} \notag \\ 
& \qquad \qquad \qquad \qquad -\widetilde{k}_{(D-d-1)}\,U_{-}^{2(d+1)/(d-1)}, \label{a19}
\end{align}
where the second equation represents the total energy conservation which can be derived by 
the variation of $n(t)$ in the action (\ref{a9}).

We get equations of motion by the variation with respect to $U_+$ and $U_-$ as follows
\begin{align}
&-\frac{4}{(d-1)^{2}}\,\ddot{U}_{+}-k_{(d)}\,U_{+} \notag \\
& \qquad \qquad -\widetilde{k}_{(D-d-1)}\,\frac{2(d+1)}{d-1}\,U_{-}^{(d+3)/(d-1)}=0, \label{a20} \\
&-\frac{4}{(d-1)^{2}}\,\ddot{U}_{-}-k_{(d)}\,U_{-}=0. \label{a21}
\end{align}
Because of the equation (\ref{a21}), we can easily find the following conserved quantities;
\begin{equation}
\epsilon =\frac{2}{(d-1)^2} \dot{U}_{-}^2+\frac{k_{(d)}}{2}{U}_{-}^2.\label{cons2}
\end{equation}
Since the system has two conserved constants (\ref{a19}) and (\ref{cons2}) for two dynamical variables, the total system is classically integrable.

In \cite{bio26}-\cite{bio28}, the integrability was discussed from the Hamiltonian viewpoint. 
It is essential idea that they looked not only for functions Poisson-commuting with the Hamiltonian $H$, but also for 
a function $F$ satisfying an equation of the form 
\begin{equation}
\{ F,\,H \}_{\text{P.B.}}=\phi H \notag 
\end{equation}
for some unknown function $\phi$. 
The Hamiltonian constraint $H=0$ indicates $\{ F,\,H \}_{\text{P.B.}}=\phi H=0$, 
therefore the function $F$ becomes a conserved quantity on this Hamiltonian constraint. 
Using this method, same consequence (\ref{a17}) was obtained in \cite{bio26}-\cite{bio28}. 
In our model, $\epsilon$ satisfies $\{ \epsilon ,\, H\}_{\text{P.B.}}=0$ where we used 
canonical momenta $P_{+}=-(4/(d-1)^{2})\dot{U}_{-}$, $P_{-}=-(4/(d-1)^{2})\dot{U}_{+}$, the Hamiltonian (\ref{a19}) 
 and 
the Poisson bracket 
\begin{equation}
\{ q_{1},\,q_{2} \}_{\text{P.B.}}\equiv\sum_{i=+,-} \Biggl[ \frac{\partial q_{1}}{\partial U_{i}}\frac{\partial q_{2}}{\partial P_{i}}
-\frac{\partial q_{1}}{\partial P_{i}}\frac{\partial q_{2}}{\partial U_{i}} \Biggr]. \notag 
\end{equation} 
$\epsilon$ commutes with the Hamiltonian and then is a conserved quantity. 
This means that the system becomes integrable, because the system has two conserved quantities for two 
dynamical variables.     
If the condition (\ref{condition}) is not satisfied, $\epsilon$ dose not commute with the Hamiltonian 
derived from (\ref{a16}) and $\{ \epsilon,\,H \}_{\text{P.B.}}\neq \phi H$. 
In the next section, we are going to show that it is quite easy to derive the analytic solutions 
by the choice of the time variable, ($M_{1}=M_{2}=1$).  

We can impose the other type of requirement that the non-linear term in the effective action (\ref{a16}) does not depend on $U_+$. It turns out that interchanging $U_+$ and $U_-$ can be achieved by the replacement $d \rightarrow D-d-1$ which means that the interchange of the internal and the external space.  This symmetry should be present because we are just considering the evolution of two homogeneous spaces. At the level of the effective action this equivalence results from the re-parametrization of the time coordinate. 
For instance, we will take $-1+\frac{M_{1}+1}{d-1}\,\Bigl[\, d-\frac{D-2}{
(d-1)X} \,\Bigr]=-1+\frac{M_{2}+1}{d-1}\,\Bigl[\, d+\frac{D-2}{
(d-1)X} \,\Bigr]=1$ in (\ref{a12}) and impose $-1+\frac{2(d-1)}{d-(D-2)/(d-1)X}=0$ which gives 
\begin{equation}
\begin{cases}
D=10,\qquad d=6 \\
D=11,\qquad d=5,\,8
\end{cases} \label{a22}
\end{equation} 
where we have used (\ref{a10}).  The  effective Lagrangian is identical to (\ref{a18}) just by interchanging the two spaces. 

\section{General Solutions}
In the previous section, we have seen that the Einstein equations for two homogeneous spaces are integrable if the total dimensions are ten or eleven. In this section, we will discuss the case $D=10,\,d=3$ which is most relevant for four-dimensional physics. 
General solutions of ($D=11,\, d=2$) and ($D=11,\, d=5$) are shown in Appendix \ref{Appendix}.

In this case, the equations of motion (\ref{a20})-(\ref{a21}) and the Hamiltonian constraint (\ref{a19}) are written by
\begin{align}
&\ddot{U}_{+}+k_{(3)}\,U_{+}
+4\,\widetilde{k}_{(6)}\,U_{-}^{3}=0, \label{b1} \\
&\ddot{U}_{-}+k_{(3)}\,U_{-}=0, \label{b2} \\
&\dot{U}_{+}\dot{U}_{-} 
+k_{(3)}\,U_{+}U_{-} 
+\widetilde{k}_{(6)}\,U_{-}^{4}=0, \label{b3}
\end{align} 
where $\widetilde{k}_{(6)}=5\,k'_{(6)}$. Let us first consider the equation of motion (\ref{b2}). 
The solution of (\ref{b2}) can be easily obtained as
\begin{equation}
U_{-}=
\begin{cases}
A_{1}\,\cos [\,\sqrt{k_{(3)}}\,t+A_{2}\, ], \qquad ~~\,(k_{(3)}>0)	\\
A_{1}\,t+A_{2}, \qquad \qquad \qquad ~~~~\,~\,(k_{(3)}=0) 			\\
A_{1}\,\cosh [\, \sqrt{-k_{(3)}}\,t+A_{2}\, ], \quad ~\,(k_{(3)}<0)
\end{cases} \label{b4}
\end{equation}
where $A_{1}$ and $A_{2}$ are constants of integrations. These equations show that the behavior of $U_{-}$ 
is controlled by the curvature of the three-dimensional Einstein space. Substituting this $U_{-}$ into the equation 
of motion (\ref{b1}), we obtain following equations of motion, 
\begin{equation}
\begin{split}
&k_{(3)}>0\,;\quad \ddot{U}_{+}+k_{(3)}\,U_{+} \\
&\qquad \qquad \quad  +4\,\widetilde{k}_{(6)}\,\Bigl(\, A_{1}\,\cos [\,\sqrt{k_{(3)}}\,t+A_{2}\, ]  \,\Bigr)^{3}=0,  \\
&k_{(3)}=0\,;\quad \ddot{U}_{+}
+4\,\widetilde{k}_{(6)}\,\Bigl(\, A_{1}\,t+A_{2}\,   \,\Bigr)^{3}=0,  \\
&k_{(3)}<0\,;\quad \ddot{U}_{+}+k_{(3)}\,U_{+} \\
& \qquad \qquad  +4\,\widetilde{k}_{(6)}\,\Bigl(\, A_{1}\,\cosh [\,\sqrt{-k_{(3)}}\,t+A_{2}\, ]  \,\Bigr)^{3}=0.  
\end{split} \label{b5}
\end{equation}
These equations of motion concretely show that the $U_{+}$ have received the forced power from the $U_{-}$. 
The internal space gives the effect of the forced oscillation to the motion of $U_{+}$.

In the cases of $k_{(3)}<0$ or $k_{(3)}>0$, it is useful to adopt the following transformations;
\begin{equation}
\begin{split}
&k_{(3)}>0\,;\quad U_{+}=f(t) \\
  &\qquad\qquad\qquad\qquad-\frac{3\,\widetilde{k}_{(6)}A_{1}^{3}}{2\,\sqrt{k_{(3)}}}\,t\, \sin [\,\sqrt{k_{(3)}}\,t+A_{2}\,] \\
  &\qquad\qquad\qquad\qquad
     +\frac{\widetilde{k}_{(6)}A^{3}_{1}}{8\,k_{(3)}}\,\cos [\,3(\,\sqrt{k_{(3)}}\,\,t+A_{2}\,)\,] \\
&k_{(3)}<0\,;\quad U_{+}=f(t) \\
  &\qquad\qquad\qquad\qquad-\frac{3\,\widetilde{k}_{(6)}A_{1}^{3}}{2\,\sqrt{-k_{(3)}}}\,t\, \sinh [\,\sqrt{-k_{(3)}}\,t+A_{2}\,] \\
  &\qquad\qquad\qquad\qquad
     +\frac{\widetilde{k}_{(6)}A^{3}_{1}}{8\,k_{(3)}}\,\cosh [\,3(\,\sqrt{-k_{(3)}}\,\,t+A_{2}\,)\,] 
\end{split} \label{b6}
\end{equation}
Using these relations,  we can simplify the equation of motion (\ref{b5}) as
\begin{equation}
\begin{split}
& \ddot{f}(t)+ k_{(3)}\,f(t)=0, \qquad ~~~~\,\,(\,k_{(3)}>0\,)\\
& \ddot{f}(t)+ (\,-k_{(3)}\,)\,f(t)=0. \qquad (\,k_{(3)}<0\,)
\end{split} \notag 
\end{equation}
Then, answers for this equation are simply 
\begin{equation}
\begin{split}
&f(t)=B_{1}\,\cos[\, \sqrt{k_{(3)}}\,\,t+B_{2} \,], (k_{(3)}>0) \\
&f(t)=B_{1}\,\sinh [\, \sqrt{-k_{(3)}}\,\,t+B_{2} \,], (k_{(3)}<0)
\end{split} \notag 
\end{equation}
 where $B_{1}$ and $B_{2}$ are the constants of integrations. 
Combining all these things, we finally obtain the solutions given by 
\begin{equation}
\begin{split}
&k_{(3)}>0\,;\quad U_{+}=B_{1}\,\cos [\, \sqrt{k_{(3)}}\,\,t+B_{2} \,] \\
&\qquad\qquad\qquad\qquad-\frac{3\,\widetilde{k}_{(6)}A_{1}^{3}}{2\,\sqrt{k_{(3)}}}\,t\, \sin [\,\sqrt{k_{(3)}}\,t+A_{2}\,] \\
  &\qquad\qquad\qquad\qquad 
     +\frac{\widetilde{k}_{(6)}A^{3}_{1}}{8\,k_{(3)}}\,\cos [\,3(\,\sqrt{k_{(3)}}\,\,t+A_{2}\,)\,], \\[6pt]
&k_{(3)}=0\,;\quad U_{+}=B_{1}\,t+B_{2}-\frac{\widetilde{k}_{(6)}}{5A^{2}_{1}}\,(A_{1}\,t+A_{2})^{5}, \\[7pt]
&k_{(3)}<0\,;\quad U_{+}=B_{1}\,\sinh [\, \sqrt{-k_{(3)}}\,\,t+B_{2} \,] \\
&\qquad\qquad\qquad\quad-\frac{3\,\widetilde{k}_{(6)}A_{1}^{3}}{2\,\sqrt{-k_{(3)}}}\,t\, \sinh [\,\sqrt{-k_{(3)}}\,t+A_{2}\,] \\
  &\qquad\qquad\qquad\quad
     +\frac{\widetilde{k}_{(6)}A^{3}_{1}}{8\,k_{(3)}}\,\cosh [\,3(\,\sqrt{-k_{(3)}}\,\,t+A_{2}\,)\,]. 
\end{split} \label{b7}
\end{equation}
The term proportional to $\widetilde{k}_{(6)}$ indicates the resonance, as the frequency of the harmonic and forced oscillation are identical.

The Hamiltonian constraint (\ref{b3}) gives a constraint on four constants of integrations, 
\begin{equation}
\begin{split}
&k_{(3)}>0\,;\quad \frac{9\widetilde{k}_{(6)}A^{3}_{1}}{8}+k_{(3)}\,B_{1}\,\cos [A_{2}-B_{2}]=0, \\
&k_{(3)}=0\,;\quad B_{1}=0, \\
&k_{(3)}<0\,;\quad \frac{9\widetilde{k}_{(6)}A^{3}_{1}}{8}+(-k_{(3)})\,B_{1}\,\sinh [A_{2}-B_{2}]=0.
\end{split} \label{b8}
\end{equation}
We shall consider the metric which is given by 
\begin{equation}
e^{2n}=e^{2A}=U_{+}U_{-},\qquad e^{2\alpha}=\sqrt{\frac{U_{+}}{U_{-}}}, \label{b9}
\end{equation}
where we have used (\ref{a10}), (\ref{a14}) and $M_{1}=M_{2}=1$. These equations shows that 
the four-dimensional part is the conformal metric, $e^{2n}=e^{2A}$. Therefore, in the four-dimensional Einstein frame, 
ten-dimensional metric is 
\begin{align}
ds_{10}^{2}&=e^{-6\alpha}[\,e^{2A}(-dt^{2}+g_{ij}(x)\,dx^{i}dx^{j})\,] \notag \\
& \qquad \qquad \qquad \qquad +e^{2\alpha}\,g_{AB}(y)\,dy^{A}dy^{B}  \notag \\
&= \Bigg( \frac{U_{+}}{U_{-}} \Biggr)^{-3/2}\,\Bigl[ U_{+}U_{-}\,(-dt^{2}+g_{ij}(x)\,dx^{i}dx^{j}) \Bigr] \notag \\
&\qquad \qquad \qquad +\Biggl( \frac{U_{+}}{U_{-}}\Biggr)^{1/2}\,g_{AB}(y)\,dy^{A}dy^{B}.  \label{b10}
\end{align}
For $\widetilde{k}_{(6)}\neq 0$, metric components are
\begin{align}
&(\,k_{(3)}>0 \,) \notag \\  
& e^{2A}=\frac{9\widetilde{k}_{(6)}A^{4}_{1}}{8\,k_{(3)}}\,\Biggl[ 
-\frac{\cos [\, \sqrt{k_{(3)}}\,\,t+B_{2} \,]}{\cos[A_{2}-B_{2}]}   \notag \\  
& -\frac{4\sqrt{k_{(3)}}}{3}\,\,t\, \sin [\,\sqrt{Y_{(3)}}\,t+A_{2}\,] \notag \\ 
& +\frac{1}{9}\,\cos [\,3(\,\sqrt{k_{(3)}}\,\,t+A_{2}\,)\,]  \Biggr]
\,\cos [\,\sqrt{k_{(3)}}\,t+A_{2}\, ], \notag \\[10pt] 
&e^{2\alpha}=\sqrt{\frac{9\widetilde{k}_{(6)}A^{2}_{1}}{8\,k_{(3)}}}\,\Biggl\{ \Biggl[ 
-\frac{\cos [\, \sqrt{k_{(3)}}\,\,t+B_{2} \,]}{\cos[A_{2}-B_{2}]}  \notag  \\  
& -\frac{4\sqrt{k_{(3)}}}{3}\,\,t\, \sin [\,\sqrt{k_{(3)}}\,t+A_{2}\,] \notag \\   
 & +\frac{1}{9}\,\cos [\,3(\,\sqrt{k_{(3)}}\,\,t+A_{2}\,)\,] \Biggr]
\,\frac{1}{\cos [\,\sqrt{k_{(3)}}\,t+A_{2}\, ] } \Biggr\}^{\frac{1}{2}},  \notag 
\end{align}
\begin{align}
&(\,k_{(3)}=0\,) \notag \\
& e^{2A}=\Biggl[ -\frac{\widetilde{k}_{(6)}}{5A^{2}_{1}}\,(A_{1}\,t+A_{2})^{5}+B_{2} \Biggr]\,(A_{1}\,t+A_{2}) , \notag \\[8pt] 
& e^{2\alpha}=\Biggl\{ \Biggl[ -\frac{\widetilde{k}_{(6)}}{5A^{2}_{1}}\,(A_{1}\,t+A_{2})^{5}
+B_{2} \Biggr]  \frac{1}{A_{1}\,t+A_{2}}  \Biggr\}^{\frac{1}{2}}, \notag 
\end{align}
\begin{align}
&(\,k_{(3)}<0 \,) \notag \\
& e^{2A}=\frac{-9\widetilde{k}_{(6)}A^{4}_{1}}{8\,(-k_{(3)})}\,\Biggl[ 
\frac{\sinh [\, \sqrt{-k_{(3)}}\,\,t+B_{2} \,]}{\sinh[A_{2}-B_{2}]}    \notag \\
&+\frac{4\sqrt{-k_{(3)}}}{3}\,\,t\, \sinh [\,\sqrt{-k_{(3)}}\,t+A_{2}\,] \notag \\  
&+\frac{1}{9}\cosh [\,3(\,\sqrt{-k_{(3)}}\,\,t+A_{2}\,)\,]  \Biggr]
\cosh [\,\sqrt{-k_{(3)}}\,t+A_{2}\, ], \notag \\[10pt]
& e^{2\alpha}=\sqrt{\frac{-9\widetilde{k}_{(6)}A^{2}_{1}}{8\,(-k_{(3)})}}\,\Biggl\{ \Biggl[ 
\frac{\sinh [\, \sqrt{-k_{(3)}}\,\,t+B_{2} \,]}{\sinh[A_{2}-B_{2}]}   \,   \notag \\ 
& +\frac{4\sqrt{-k_{(3)}}}{3}\,\,t\, \sinh [\,\sqrt{-k_{(3)}}\,t+A_{2}\,]  \notag \\  
  & +\frac{1}{9}\cosh [3(\,\sqrt{-k_{(3)}}\,\,t+A_{2}\,)\,] \Biggr]
\,\frac{1}{\cosh [\sqrt{-k_{(3)}}\,t+A_{2} ] } \Biggr\}^{\frac{1}{2}}.
 \label{b11}
\end{align}

\section{Cosmological Characteristic for (D=10, d=3)}

The solutions that we have obtained in the previous section include a metric which has realistic dimensions 
 $D=10,\,d=3$. The total dimension of this spacetime is ten dimension equal to the critical dimension of superstrings and the physical space-time has four dimensions.
Therefore, it is very interesting how 
the universe evolve with time. In this section, we shall consider the behavior of
 the ten-dimensional spacetime. We will analytically show  that the four-dimensional part of the ten-dimensional spacetime 
accelerates eternally, which has been analyzed in \cite{bio5} by the qualitative method and in \cite{bio29} by the fixed point analysis on the phase space. 

We shall consider metric (\ref{b11}). In $k_{(3)}>0$, $e^{2\,A}$ and $e^{2\,\alpha}$ take oscillatory behavior. 
$e^{2\,A}$ starts from zero and end up with zero because $U_{-}=A_{1}\,\cos[\sqrt{k_{(3)}}\,t+A_{1}]$ oscillates between two zeros. 
On the other hand, $e^{2\,\alpha}$ diverges when $U_{-}$ and $e^{2\,A}$ become zero, and then, the case of $k_{(3)}>0$ 
may not have a stable internal space. 

Similarly, in $k_{(3)}=0$, the scale factor of the internal space 
diverges at $U_{-}=A_{1}\,t+A_{2}=0$ and $e^{2\,A}$ takes zero when $U_{-}=0$. 
For $k_{(3)}=0$ and $\widetilde{k}_{6}<0$,  the asymptotic behavior becomes 
$e^{2\,A}\rightarrow t^{6}$ and $e^{2\,\alpha}\rightarrow t^{2}$ at $t\rightarrow \infty$ and 
the ten-dimensional metric (\ref{b10}) has the behavior as follows
\begin{equation}
ds_{10}^{2}\rightarrow (\,-dt^{2}+ds_{3}^{2}\,)+t^{2}\,ds_{6}^{2}.  \label{c1}
\end{equation}

This means that the four-dimensional part of the metric does not depend on $t$ in the ten-dimensional frame. 
If we take $ds^{2}_{3}$ to the three-dimensional Euclid space, the four-dimensional part becomes the
Minkowski spacetime at $t \rightarrow \infty$. The internal space becomes large in this region. 

If $k_{(3)}<0$, an interesting phenomenon occurs. The three-dimensional space expands with acceleration eternally. This acceleration is extracted from the negative curvature of the 
internal space, $\widetilde{k}_{(6)}<0$.  The curvature of the internal space acts like a positive cosmological constant in four dimensions. 
We assume the internal space is the Einstein space with the negative curvature. 
This case, $k_{(3)}<0$ and $\widetilde{k}_{(6)}<0$, is equivalent to the situation suggested in \cite{bio5} in which 
it was shown by scalar perturbations of scale factors 
that the acceleration of the three-dimensional space occurs in the four-dimensional Einstein frame. 
In \cite{bio29} the fixed point analysis also indicated that the eternal acceleration occurs for $k_{(3)}<0$ and $\widetilde{k}_{(6)}<0$. 
We can confirm these facts by using the analytic solutions.

\begin{figure}[h]
  \begin{center}
   \includegraphics[width=54mm]{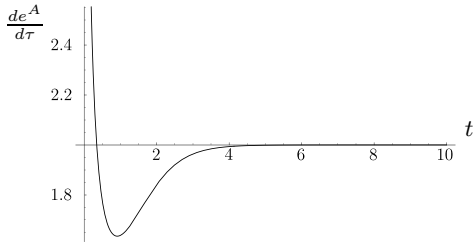}
	   \begin{picture}(0,109)
		   \put(-174,80){$\frac{de^{A}}{d\tau}$}
		   \put(0,40){$t$}
	   \end{picture}
  \end{center}
  \caption{The velocity of $e^{A}$. $k_{(3)}=-1$, $A_{2}=0.5$, $B_{2}=0$.}
  \label{fig1}
 \end{figure}

\begin{figure}[h]
  \begin{center}
   \includegraphics[width=55mm]{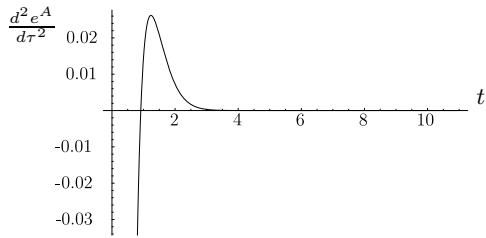}
	   \begin{picture}(21,0)
		   \put(-178,77){$\frac{d^{2}e^{A}}{d\tau^{2}}$}
		   \put(0,50){$t$}
	   \end{picture}
  \end{center}
  \caption{The acceleration of $e^{A}$. $k_{(3)}=-1$, $A_{2}=0.5$, $B_{2}=0$.}
  \label{fig2}
\end{figure}

Defining a proper time 
$d\tau \equiv \sqrt{U_{+}\,U_{-}}\,dt$, the velocity and acceleration of $e^{A}$ are given by 
$de^{A}/d\tau$ and $d^{2}e^{A}/d\tau^{2}$. These velocity and acceleration are shown in fig.[\ref{fig1}]-[\ref{fig2}] where we 
neglect overall factor in (\ref{b11}).
From these figures, we can extract following facts. 
The three-dimensional space evolves with the very large positive velocity and 
negative acceleration, then the three-dimensional space decelerates quickly at a first stage. The acceleration of the three-dimensional 
space turns to positive at some time, and then the acceleration decreases gradually. The three-dimensional space 
finally expands with a positive velocity and the zero acceleration at the infinite future. 
In particular, the three-dimensional space accelerates forever. This fact coincides with \cite{bio5} and \cite{bio29}. 

We shall consider the asymptotic behavior of $e^{2\,A}$ and $e^{2\,\alpha}$ in (\ref{b11}). 
$U_{-}=A_{1}\,\cosh[\sqrt{-k_{(3)}}\,t+A_{2}]$ is not singular for appropriate constants of integrations. $A_{1}$ and $A_{2}$, 
which implies $e^{2\,\alpha}$ may not diverge when $e^{2\,A} \rightarrow 0$. 
For $t \rightarrow \infty$, $e^{2\,A}\rightarrow e^{4\,\sqrt{-k_{(3)}}\,t}$,
$e^{2\,\alpha}\rightarrow e^{\,\sqrt{-k_{(3)}}\,t}$ and 
\begin{equation}
ds^{2}_{10}\rightarrow e^{\,\sqrt{-k_{(3)}}\,t}(\,-dt^{2}+ds^{2}_{3}+ds^{2}_{6}\,). \label{c2}
\end{equation}
It is found that this ten-dimensional metric is conformally equivalent to $-dt^{2}+ds^{2}_{3}+ds^{2}_{6}$ at large $t$. 
In four-dimensional Einstein frame, this metric (\ref{c2}) behaves as 
\begin{equation}
\begin{split}
ds^{2}_{10}\rightarrow &e^{\,-3\sqrt{-k_{(3)}}\,t}[e^{\,4\sqrt{-k_{(3)}}\,t}(-dt^{2}+ds^{2}_{3})] \\
& \qquad \qquad \qquad \qquad \qquad \quad +e^{\,\sqrt{-k_{(3)}}\,t}ds^{2}_{6}. \label{c3}
\end{split}
\end{equation}
The above metric (\ref{c3}) shows that the internal space also expands.  
 The proper time, for the four-dimensional frame, is given by
$\tau \sim e^{\,2\sqrt{-k_{(3)}}\,t}$ and then, the three-dimensional space expands with $e^{2\,A}\sim \tau^{2}$ which 
shows that the expansion has the uniform velocity at large $\tau$. The three-dimensional space has 
a negative curvature $k_{(3)}<0$ in this case.  Therefore the four-dimensional part can be the Milne universe for the 
four-dimensional frame. The metric (\ref{c3}) also indicates that the internal space becomes 
large with $e^{2\alpha} \sim e^{\,\sqrt{-k_{(3)}}\,t}$, and then it is intuitively expected that 
the curvature of the internal space to 
decrease. In fact, it is found in (\ref{a9}) that the potential term 
$ -\widetilde{k}_{(6)}e^{-8\alpha} $ vanishes at $t \rightarrow \infty$.

If $t \rightarrow 0$, $ds^{2}_{10}\rightarrow t^{-1/2}(-dt^{2}+ds^{2}_{3})+t^{1/2}ds^{2}_{6}$ and in 
the four-dimensional frame, 
\begin{equation}
ds^{2}_{10}\rightarrow t^{-3/2}[\,t\,(-dt^{2}+ds^{2}_{3})\,]+t^{1/2}ds^{2}_{6}. \label{c4}
\end{equation}
The proper time for the four-dimensional frame is given by $\tau \sim t^{3/2}$ and the three-dimensional space 
expands with $e^{2\,A}\sim \tau^{2/3}$. The acceleration of this scale factor is  
$d^{2} e^{A}/d\tau^{2} \sim -(2/9)\, \tau^{-5/3}$ and then the acceleration diverges at $\tau\rightarrow 0$.

As a final example, we shall consider $\widetilde{k}_{(6)}=0$ and $k_{(3)}<0$. In this case, we can find that the constants of integrations satisfy $A_{2}=B_{2}$ in (\ref{b8}). Using (\ref{b4}), (\ref{b7}) and (\ref{b9}), 
the asymptotic behavior of $e^{2\,A}$ and $e^{2\,\alpha}$  
are $e^{2\,\alpha}\rightarrow \text{const.}$ and $e^{2\,A}\rightarrow e^{2\,\sqrt{-k_{(3)}}\,t}$ at 
$t \rightarrow \infty$. The ten-dimensional metric leads to
\begin{equation}
ds^{2}_{10}\rightarrow e^{2\,\sqrt{-k_{(3)}}\, t} (-dt^{2}+ds^{2}_{3})+ds^{2}_{6}. \label{c5}
\end{equation}
This metric shows that the internal space does not depend on $t$.
The proper time is defined as $\tau \sim e^{\,\sqrt{-k_{(3)}}\,t}$ in ten dimension. Therefore, the ten-dimensional metric is represented as 
\begin{equation}
ds^{2}_{10}\rightarrow (-d\tau^{2}+\tau^{2}\, ds^{2}_{3})+ds^{2}_{6} \label{c6}
\end{equation}
whose structure is the product space of the Milne universe and a flat six-dimensional space. 
It is possible to transform the Milne universe into the Minkowski spacetime by coordinate transformations.
In this case, the above metric (\ref{c6}) becomes the product spacetime with the four-dimensional Minkowski spacetime and 
the flat six-dimensional internal space at $t\rightarrow \infty$.

\section{Conclusions}

We have considered the vacuum Einstein equations in the $D$-dimensional spacetime and 
obtained integrable cosmological models.
Those Einstein equations have two potential terms arising from the curvature of the $d$- and $(D-d-1)$-dimensional 
Einstein spaces. It was thought that solving Einstein equations with two curved spaces are very difficult. 
However we have pointed out that the total dimension should be $D=10$ or $D=11$ 
to make  those Einstein equations integrable as cosmological models.  
The integrability is guaranteed by the conserved quantity which commutes with the Hamiltonian. 
The integrable system does not necessarily have the analytic solutions. 
It is very important to note that the time variable used in this paper 
easily realizes the analytic solutions.   
It is interesting that models with superstrings or M-theory are more tractable as cosmological 
models than other dimensional models.

For ($D=10$, $d=3$), we have obtained the accelerating universe with two spatial parts whose curvature is negative. 
The three-dimensional space expands with the acceleration, but the six-dimensional 
internal space also expands. The external space finally approaches the expansion whose acceleration tends to zero. 
It may be difficult for this case to give an account of the realistic acceleration at late time.  
To obtain more realistic models, we need to construct a model whose internal space is fixed dynamically whereas our space is going to expand more drastically. In the context of pure gravity solutions we have treated in this paper, we cannot get such solutions. 
The flux field, dilaton and the world volume actions may have a role for the interesting behavior such as fixing the internal spaces \cite{bio23}. 

It would be more interesting to find solutions for more general setup.  

\begin{acknowledgments}
We would like to thank N. D. Hari Dass for valuable discussions and Eiichi Takasugi and Tetsuyuki Yukawa for useful comments. 
We would like to thank Nobuyoshi Ohta for informing us papers \cite{bio26}-\cite{bio29}. 
\end{acknowledgments}

\appendix

\section{General solutions in $D=11$}\label{Appendix}

We can obtain other solutions corresponding to ($D=11,\,d=2$) and ($D=11,\,d=5$) from the equations 
(\ref{a18}) and (\ref{a19}).

 In ($D=11,\,d=2$), the equations of motion and the Hamiltonian constraint are 
\begin{align}
&4\,\ddot{U}_{+}+k_{(2)}\,U_{+}
+6\,\widetilde{k}_{(8)}\,U_{-}^{5}=0, \label{b12} \\
&4\ddot{U}_{-}+k_{(2)}\,U_{-}=0, \label{b13} \\
&4\dot{U}_{+}\dot{U}_{-} 
+k_{(2)}\,U_{+}U_{-} 
+\widetilde{k}_{(8)}\,U_{-}^{6}=0, \label{b14}
\end{align}
and for ($D=11,\,d=5$), 
\begin{align}
&\frac{1}{4}\,\ddot{U}_{+}+k_{(5)}\,U_{+}
+3\,\widetilde{k}_{(5)}\,U_{-}^{2}=0, \label{b15} \\
&\frac{1}{4}\,\ddot{U}_{-}+k_{(5)}\,U_{-}=0, \label{b16} \\
&\frac{1}{4}\,\dot{U}_{+}\dot{U}_{-} 
+k_{(5)}\,U_{+}U_{-} 
+\widetilde{k}_{(5)}\,U_{-}^{3}=0. \label{b17}
\end{align}
For ($D=11,\,d=2$), solutions are given by 
\begin{equation}
U_{-}=
\begin{cases}
A_{1}\,\cos [\,\frac{\sqrt{k_{(2)}}}{2}\,t+A_{2}\, ], \qquad ~~\,(k_{(2)}>0)	\\
A_{1}\,t+A_{2}, \qquad \qquad \qquad ~~~~\,~(k_{(2)}=0) 			\\
A_{1}\,\cosh [\, \frac{\sqrt{-k_{(2)}}}{2}\,t+A_{2}\, ], \quad ~\,\,(k_{(2)}<0)
\end{cases} \label{b18}
\end{equation}
\begin{align}
&k_{(2)}>0\,;\quad U_{+}=B_{1}\,\cos [\, \frac{\sqrt{k_{(2)}}}{2}\,\,t+B_{2} \,] \notag \\
&\qquad\qquad\qquad -\frac{15\,\widetilde{k}_{(8)}A_{1}^{5}}{16\,\sqrt{k_{(2)}}}\,\,t\, \sin [\,\frac{\sqrt{k_{(2)}}}{2}\,t+A_{2}\,] \notag \\
  &\qquad\qquad\qquad
     +\frac{15\widetilde{k}_{(8)}A^{5}_{1}}{64\,k_{(2)}}\,\cos [\,3(\,\frac{\sqrt{k_{(2)}}}{2}\,\,t+A_{2}\,)\,] \notag \\
     &\qquad\qquad\qquad+\frac{\widetilde{k}_{(8)}A^{5}_{1}}{64\,k_{(2)}}\,\cos [\,5(\,\frac{\sqrt{k_{(2)}}}{2}\,\,t+A_{2}\,)\,], \notag 
\end{align}
\begin{align}
&k_{(2)}=0\,;\quad U_{+}=B_{1}\,t+B_{2}-\frac{\widetilde{k}_{(8)}}{28A^{2}_{1}}\,(A_{1}\,t+A_{2})^{7}, \notag
\end{align}
\begin{align}
&k_{(2)}<0\,;\quad U_{+}=B_{1}\,\sinh [\, \frac{\sqrt{-k_{(2)}}}{2}\,\,t+B_{2} \,] \notag \\
&\qquad\qquad\qquad-\frac{15\,\widetilde{k}_{(8)}A_{1}^{5}}{16\,\sqrt{-k_{(2)}}}\,\,t\, \sinh [\,\frac{\sqrt{-k_{(2)}}}{2}\,t+A_{1}\,] \notag \\
  &\qquad\qquad\qquad
     +\frac{15\widetilde{k}_{(8)}A^{5}_{1}}{64\,k_{(2)}}\,\cosh [\,3(\,\frac{\sqrt{-k_{(2)}}}{2}\,\,t+A_{2}\,)\,] \notag \\
     &\qquad\qquad\qquad+\frac{\widetilde{k}_{(8)}A^{5}_{1}}{64\,k_{(2)}}\,\cosh [\,5(\,\frac{\sqrt{-k_{(2)}}}{2}\,\,t+A_{2}\,)\,], 
 \label{b19}
\end{align}
and constraints are given by 
\begin{align}
&k_{(2)}>0\,;\quad \frac{5\widetilde{k}_{(8)}A^{5}_{1}}{4}+k_{(2)}\,B_{1}\,\cos [A_{2}-B_{2}]=0, \notag \\
&k_{(2)}=0\,;\quad B_{1}=0,  \\
&k_{(2)}<0\,;\quad \frac{5\widetilde{k}_{(8)}A^{5}_{1}}{4}+(-k_{(2)})\,B_{1}\,\sinh [A_{2}-B_{2}]=0. \notag 
 \label{b20}
\end{align}
For ($D=11,\,d=5$), solutions are
\begin{equation}
U_{-}=
\begin{cases}
A_{1}\,\cos [\,2\sqrt{k_{(5)}}\,t+A_{2}\, ], \qquad ~~\,\,\,(k_{(5)}>0)	\\
A_{1}\,t+A_{2}, \qquad \qquad \qquad ~~~~~~\,~\,\,(k_{(5)}=0) 			\\
A_{1}\,\cosh [\, 2\sqrt{-k_{(5)}}\,t+A_{2}\, ], \quad ~~\,(k_{(5}<0)
\end{cases} \label{b21}
\end{equation}
\begin{align}
&k_{(5)}>0\,;\quad U_{+}=B_{1}\,\cos [\, 2\sqrt{k_{(5)}}\,\,t+B_{2} \,] \notag \\
&\qquad\qquad \qquad \quad -\frac{3\,\widetilde{k}_{(5)}A_{1}^{2}}{2\,k_{(5)}} \notag \\
  &\qquad\qquad \qquad \quad+\frac{\widetilde{k}_{(5)}A^{2}_{1}}{2\,k_{(5)}}\,\cos [\,2(\,2\sqrt{k_{(5)}}\,\,t+A_{2}\,)\,], \notag 
\end{align}
\begin{align}
&k_{(5)}=0\,;\quad U_{+}=B_{1}\,t+B_{2}-\frac{\widetilde{k}_{(5)}}{A_{1}^{2}}\,(A_{1}\,t+A_{2})^{4}, \notag 
\end{align}
\begin{align}
&k_{(5)}<0\,;\quad U_{+}=B_{1}\,\sinh [\, 2\sqrt{-k_{(5)}}\,\,t+B_{2} \,] \notag \\
&\qquad\qquad \qquad \quad-\frac{3\,\widetilde{k}_{(5)}A_{1}^{2}}{2\,k_{(5)}} \notag \\
  &\qquad\qquad\qquad\quad
     +\frac{\widetilde{k}_{(5)}A^{2}_{1}}{2\,k_{(5)}}\,\cosh [\,2(\,2\sqrt{-k_{(5)}}\,\,t+A_{2}\,)\,]. 
\label{b22}
\end{align}
The constraints are 
\begin{equation}
\begin{split}
&k_{(5)}>0\,;\quad k_{(5)}\,A_{1}\,B_{1}\,\cos [A_{2}-B_{2}]=0, \\
&k_{(5)}=0\,;\quad B_{1}=0, \\
&k_{(5)}<0\,;\quad (-k_{(5)})\,A_{1}\,B_{1}\,\sinh [A_{2}-B_{2}]=0.
\end{split} \label{b23}
\end{equation}
For ($D=11,\,d=2$) and three dimensional Einstein frame, the metric is 
\begin{align}
ds_{11}^{2}&=\Bigg( \frac{U_{+}}{U_{-}} \Biggr)^{-8/3}\,\Bigl[ (U_{+}U_{-})^{2}\,(-dt^{2}+g_{ij}(x)\,dx^{i}dx^{j}) \Bigr] \notag \\
&\qquad \qquad \qquad \qquad +\Biggl( \frac{U_{+}}{U_{-}}\Biggr)^{1/3}\,g_{AB}(y)\,dy^{A}dy^{B},  \label{b24} 
\end{align}
\begin{align}
& e^{2\,A}=
e^{2\,n}=(U_{+}\,U_{-})^{2},\qquad e^{2\,\alpha}=\Biggl( \frac{U_{+}}{U_{-}}\Biggr)^{1/3}. \label{b25}
\end{align}
For ($D=11,\,d=5$) and six dimensional Einstein frame, the metric is given by 
\begin{align}
ds_{11}^{2}&=\Bigg( \frac{U_{+}}{U_{-}} \Biggr)^{-5/6}\Bigl[ (U_{+}U_{-})^{1/2}(-dt^{2}+g_{ij}(x)\,dx^{i}dx^{j}) \Bigr] \notag \\
& \qquad \qquad \qquad \qquad +\Biggl( \frac{U_{+}}{U_{-}}\Biggr)^{2/3}g_{AB}(y)\,dy^{A}dy^{B},  \label{b26} 
\end{align}
\begin{align}
& e^{2\,A}
=e^{2\,n}=(U_{+}\,U_{-})^{1/2},\qquad e^{2\,\alpha}=\Biggl( \frac{U_{+}}{U_{-}}\Biggr)^{2/3}. \label{b27}
\end{align}

\end{document}